\documentclass[aps,epsfigure,twocolumn,superscriptaddress,showkeys,nofootinbib]{revtex4-1}
\usepackage[colorlinks=true,linkcolor=blue,urlcolor=blue,citecolor=blue,pdfusetitle]{hyperref}
\usepackage[utf8]{inputenc}
\usepackage[english]{babel}
\usepackage{amsmath}
\usepackage[caption = false]{subfig}
\usepackage{graphicx,epstopdf}
\usepackage{blindtext}
\usepackage[table,xcdraw]{xcolor}
\usepackage{lipsum}
\usepackage{amsfonts}
\usepackage{bbm}
\usepackage{amssymb}
\usepackage{enumerate}
\usepackage{color}
\usepackage{latexsym}
\usepackage{times,txfonts}
\usepackage[normalem]{ulem}
\graphicspath{ {./figures/} }
\usepackage{subcaption}
\usepackage{amsmath}
\usepackage{algorithm}
\usepackage{algpseudocode}
\usepackage{tikz}
\usetikzlibrary{quantikz}
\usepackage{booktabs}
\usepackage{ragged2e}

\begin{document}

\title{Variational quantum simulation of a nonadditive relaxation dynamics in a qubit coupled to a finite-temperature bath}

\author{Lucas Q. Galvão}
\email{lqgalvao3@gmail.com}
\affiliation{Latin American Quantum Computing Center, SENAI CIMATEC, Salvador, Brasil.}
\affiliation{QuIIN - Quantum Industrial Innovation, Centro de Competência Embrapii Cimatec. SENAI CIMATEC, Av. Orlando Gomes, 1845, Salvador, BA, Brazil CEP 41850-010}
\affiliation{Universidade SENAI CIMATEC, Salvador, BA, Brazil}

\author{Clebson S. Cruz}
\email{clebson.cruz@ufob.edu.br}
\affiliation{Centro de Ciências Exatas e das Tecnologias, Universidade Federal do Oeste da Bahia - Campus Reitor Edgard Santos. Rua Bertioga, 892, Morada Nobre I, 47810-059 Barreiras, Bahia, Brasil.}

\author{Antonio Cesar do Prado Rosa Junior}
\email{antoniocprj@ufob.edu.br}
\affiliation{Centro de Ciências Exatas e das Tecnologias, Universidade Federal do Oeste da Bahia - Campus Reitor Edgard Santos. Rua Bertioga, 892, Morada Nobre I, 47810-059 Barreiras, Bahia, Brasil.}

\author{Marcelo A. Moret}
\email{mamoret@gmail.com}
\affiliation{Universidade SENAI CIMATEC, Salvador, BA, Brazil}

\begin{abstract}

\noindent In this paper, we present an application of the variational quantum simulation (VQS) framework to capture finite‑temperature open‑system dynamics on near‑term quantum hardware. By embedding the generalized amplitude‑damping channel into the VQS algorithm, we model energy exchange with a thermal bath through its Lindblad representation and thereby simulate realistic dissipative effects. To explore a wide range of activation behaviors, we introduce a nonadditive relaxation‑time model using a generalized form of the Arrhenius law, based on the phenomenological parameter $q$. 
We compare our method on a driven spin‑1/2 qubit subject to both static and composite time‑dependent fields, comparing population evolution and trace‑distance errors against exact solutions. Our results demonstrate that (i) VQS accurately maps the effective nonunitary generator $H_{\mathrm{eff}}$ under generalized amplitude damping, (ii) smoother drive envelopes induced by nonaddtive parameters suppress high‑frequency components and yield lower simulation errors, and (iii) the variational manifold exhibits dynamical selectivity, maintaining mapping fidelity even as the exact solution’s sensitivity to $q$ increases. 

\end{abstract}
\maketitle

\section{Introduction}

Research into quantum materials forms the basis for the second quantum revolution \cite{dowling2003quantum, GYONGYOSI201951, doi:10.1021/jacs.9b00984}. Simulating the dynamics of such systems is crucial for advancing quantum-based technologies, since it is possible to understand how quantum properties can be used to our advantage \cite{RevModPhys.86.153, Buluta_2011, PhysRevLett.82.5381, doi:10.1126/science.1177838}. In realistic scenarios, the interaction between the system and its environment must be taken into account in order to capture behaviors near practical implementations, leading to the simulation of open quantum systems \cite{lloyd1996universal, barreiro2011open, breuer2002theory}. In this scenario, quantum simulation emerges as a powerful tool with the promise of using, in principle, fewer resources for the simulation of quantum systems \cite{feynman2018simulating}.

However, the noise and qubit restrictions in current quantum computers are a limiting factor for the accuracy of quantum simulations, which led to the name of the Noisy Intermediate-Scale Quantum (NISQ) era \cite{preskill2018quantum}. Variational quantum algorithms (VQA) have emerged as a potential strategy to obtain quantum advantage on NISQ devices \cite{cerezo2021variational}. In this context, variational quantum simulation (VQS) appears as a possibility to simulate open quantum systems using shallow circuits \cite{yuan2019theory}. However, the existing implementations typically rely on Hamiltonians that neglect finite-temperature dissipative effects, assuming zero-temperature or purely unitary dynamics \cite{yuan2019theory, endo2020variational, li2017efficient, luo2024variational}. These approaches often model dissipation via simplistic Lindblad operators without explicitly capturing the qubit analogue of a thermal bath. Additionally, they do not consider direct responses of the system relaxation time to dissipative dynamics, thereby neglecting the realistic relaxation dynamics inherent to open quantum systems \cite{Jaksic1997Spectral}. 

Relaxation dynamics govern how quickly a system returns to equilibrium after perturbations, reflecting the interplay between coherent Hamiltonian motion and dissipative interactions with its surroundings \cite{Jaksic1997Spectral, wang2011quantum}. These dissipative processes can dramatically influence qubit coherence times and gate fidelity, underscoring the necessity of incorporating accurate relaxation models into simulations for reliable performance predictions \cite{yamaguchi1999crystal, hu2002decoherence, tolunay2023hamiltonian}. For its characterization, several models have been proposed to study relaxation dynamics in materials that exhibit thermal activation mechanisms \cite{Arnold1996Self-consistent}. A common approach for a wide range of materials is the so-called Arrhenius-type thermal activation with approximately temperature-independent activation energy, where the relaxation time is proportional to the exponential of the inverse temperature  \cite{prester2011slow, bochenek2013technology, Ferbinteanu2005Single-chain, Yan-Rong2009Magnetic, Sirenko2024The}.  

However, the existence of phenomena not described by this Arrhenius-type equation led to the development of alternative models in order to capture non-Arrhenius behaviors observed in experimental data \cite{demsar2020non, erdem2011nonequilibrium, blackmore2023characterisation, levit2019generalized, Liu2020Nonlinear}. Such aspects have already been observed in the diffusivity of supercooled liquids near the glass transition and in water-type extended simple point charge models \cite{geske2018molecular, huang2018interplay}, food systems \cite{Stroka01032011}, chemical reactions \cite{10.1063/1.3555763, Galamba_2017} and several biological processes \cite{NISHIYAMA2009325, doi:10.1021/acs.jpcb.7b03698}. A more general proposition was made in \cite{rosa2019characterization}, where the authors discuss a theoretical basis for the physical interpretation of the characteristic non-Arrhenius behavior of several diffusive processes, considering a nonadditive model.

In this work, we present an extension of the VQS framework to the simulation of finite‑temperature open‑system dynamics by embedding the generalized amplitude-damping channel, thereby capturing realistic energy exchange with a thermal bath. To address a broad spectrum of activation behaviors, we introduce a nonadditive relaxation‑time model that recovers the classical Arrhenius law as a special case while accommodating non‑Arrhenius regimes. Through comprehensive numerical studies on a driven qubit, we demonstrate that our approach not only faithfully maps the effective Hamiltonian dynamics but also markedly reduces simulation errors under nonadditive activation. 

The remainder of the paper is organized as follows. Section \ref{sec:2} describes the nonadditive model for a qubit subject to a finite-temperature bath, a model using generalized amplitude damping. Section \ref{sec:3} reviews the theoretical foundations of VQS. Section \ref{sec:results} presents our simulation results and error analysis. Section \ref{sec:5} concludes with a summary of key findings.

\section{Nonadditive Model for a qubit subjected to a finite temperature bath}
\label{sec:2}

One of the main frameworks used to describe the time evolution of open quantum systems, under the assumption of Markovian dynamics, is the Lindblad Master Equation (LME), written as

\begin{equation} \label{eq:LME}
\frac{d\rho}{dt} = -\frac{i}{\hbar} \left[H, \rho \right] + \sum_k J_k \left( L_k \rho L_k^\dagger - \frac{1}{2} \{ L_k^\dagger L_k, \rho \} \right),
\end{equation}
where $\rho$ is the reduced density matrix of the system, $H$ is the system Hamiltonian, $\{L_k\}$ are Lindblad operators describing the coupling of the system to the environment and $\{J_k\}$ represents the coupling strength \cite{lindblad1976generators, breuer2002theory, manzano2020short}. Here, the first term on the right-hand side of the eq. \ref{eq:LME} characterizes the von Neumann equation for a conservative system, while the second term represents the system's dissipator.

For a two-level system, its inverse temperature $\beta$ can be trivially described by the rate between both populations even for a non-equilibrium case \cite{de2020full}:

\begin{equation}
    \beta = \Delta\varepsilon^{-1} ln \left ( \frac{\rho_{00}(t)}{\rho_{11}(t)} \right ) \ ,
\end{equation}
where $\rho_{00}$ and $\rho_{11}$ are, respectively, the ground and excited state probabilities, and $\Delta \varepsilon$ is the qubit energy gap. 

As a result, the coupling of the system to the environment can directly influence its temperature in the presence of a dissipative channel capable of changing its population dynamics \cite{RevModPhys.59.1}. A direct case of this interaction is the finite-temperature bath, which is modeled considering the generalized amplitude damping channel, described by the dissipator \cite{fujiwara2004estimation}

\begin{equation} \label{eq:env}
    \begin{aligned}
    \mathcal{L}(\rho) &= J_1(1-f) \Big[\sigma_{-} \rho \sigma_{+} 
    - \frac{1}{2} \{\sigma_{+}\sigma_{-}, \rho\} \Big] \\
    &\quad + J_2 f \Big[\sigma_{+} \rho \sigma_{-} 
    - \frac{1}{2} \{\sigma_{-}\sigma_{+}, \rho\} \Big] \ .
    \end{aligned}
\end{equation}

Here, $f \in [0, 0.5]$ corresponds to the statistical distribution of the bath. For a fermion bath, $f(E) = (1+e^{(E - E_F)\beta_E})^{-1}$ corresponds to the Fermi-Dirac distribution for a probability of occupation of a quantum energy state $E$, with $E_F$ being the Fermi-Energy and $\beta_E = 1/T_{env}$ the fixed inverse temperature of the environment \cite{PhysRevB.101.155134, nesbet1961approximate, nusseler2020efficient, ghosh2012fermionic}.

The action of an environment at some fixed temperature on a qubit initially as a closed system can be described by the thermal density operator \cite{de2020full, bernardo2020unraveling}

\begin{equation} \label{eq:rho_open}
    \hat{\rho}(t) =  \begin{pmatrix}
1 - \rho_{11} (1-p(t)) - p(t) f & \rho_{01}\sqrt{1-p(t)} \\
\rho_{10} \sqrt{1-p(t)} & \rho_{11} (1-p(t)) + p(t) f \\
\end{pmatrix} \ .
\end{equation}
 
In this expression, the dissipator parameter $p$ is the decay probability associated with the excited state, given by $p = (1-e^{-\lambda t}) \in [0, 1]$, with $\lambda$ being the damping constant \cite{de2020full}. One can check $\hat\rho (t)$ satisfies the initial condition as a closed system when $t = 0$ and the equilibrium condition $\beta = \beta_E$ for $t \longrightarrow\infty$.

Fermionic systems, such as electrons, can be manipulated through the application of external magnetic fields. These fields may be static, generated by a direct current ($B_{\text{DC}}$), or time-dependent, produced by an alternating current ($B_{\text{AC}}$) and typically associated with a driving frequency of the form $\omega t$. The oscillation frequency $\omega$ plays a crucial role in the system's dynamics, particularly in scenarios where the system is able to follow the external field — that is, when the driving frequency is comparable to the system's intrinsic response or relaxation rate. As demonstrated in \cite{topping2018ac}, the interplay between $\omega$ and the system's relaxation time can give rise to distinct dynamical regimes governed by thermal activation processes:

\begin{enumerate}[(i)]
    \item \label{item:i} For the regime $\tau \gg 1/ \omega$, it corresponds to the d.c. limit, where the system responds instantaneously to the a.c. field, obtaining the equilibrium response and exchange of energy with the lattice.
    
    \item \label{item:ii} For the regime $\tau \ll 1/\omega$,  the perturbing field oscillates too quickly, preventing the magnetic moment of the system from responding and causing it to fail to equilibrate and exchange energy with the lattice.
    
    \item \label{item:iii} For the regime $\tau \approx 1/\omega$, the intermediate regime, where the oscillating magnetic field frequency is similar to the system’s magnetic relaxation timescale, provides a more complex response, such as phase lag and dissipation. 
\end{enumerate}

In most systems, it is usual to observe the predominance of thermal activation in relaxation time on these different regimes. A common approach to this characterization in quantum systems was experimentally proven \cite{prester2011slow, bochenek2013technology, Ferbinteanu2005Single-chain, Yan-Rong2009Magnetic, Sirenko2024The} to obey Arrhenius-type thermal activation with approximately temperature-independent activation energy 

\begin{equation} \label{eq:arrh}
    \tau = \tau_0 \exp{\left (E_A \beta\right )} \ ,
\end{equation}
where $E_A$ is the activation energy of the system and $\tau_0$ characterizes the relaxation time at infinite temperature, usually $\tau_0 \approx 10^{-9} \sim 10^{-11}$. Arrhenius's law implies the need for the system to overcome an energy barrier to initiate the response dynamics \cite{truhlar2001convex}. In regime \ref{item:iii}, this response is inversely proportional to the frequency of the AC field.

This model does not capture nonadditive behaviors, so different models have been proposed \cite{heitler1936time, korringa1950nuclear, demsar2020non, erdem2011nonequilibrium, blackmore2023characterisation, levit2019generalized, Liu2020Nonlinear}. Heitler-Teller's law takes into account a relaxation time directly proportional to the inverse temperature of the system, a common dependence in magnetic nuclear spins \cite{heitler1936time, korringa1950nuclear}. The Valgel-Fulcher-Tamman's law and Basser's law take into account critical temperatures in systems that typically exhibit phase transitions \cite{demsar2020non, erdem2011nonequilibrium, blackmore2023characterisation}. Other approaches propose an extension of these models, such as the generalized VFT model \cite{levit2019generalized} and a VFT nonlinear modification \cite{Liu2020Nonlinear}.

In a more specific scenario, recent studies suggested the use of nonadditive models in order to study alternative ways to describe the non-Arrhenius behavior of nonadditive stochastic systems \cite{rosa2019characterization, rosa2020non, rosa2016model}. The authors introduce a class of nonlinear Fokker-Planck equations whose solutions form a set of rapidly decreasing functions \cite{beerends2003fourier}, optimized to maximize nonadditive entropies, such as the Tsallis entropy \cite{tsallis1988possible}.  In ref. \cite{SILVA2013201}, the author provides a general approach that links deviations from Arrhenius's law to quantum tunneling and classical non-extensive statistical mechanics, offering a useful criterion to distinguish quantum effects from collective classical phenomena. To search for a common approach for these descriptions, we proposed a generalization of Eq. \eqref{eq:arrh} aiming to describe non-Arrhenius phenomena in time relaxation dynamics, considering nonadditive models, written as

\begin{equation} \label{eq:nonad}
    \tau_q = \tau_0 \left [1 + (q-1) E_A\beta \right ]^{\frac{1}{q-1}} \ ,
\end{equation}
where Arrhenius-like behavior is recovered when $q \longrightarrow 1$. In principle, the parameter $q$ can be estimated by knowing the microscopic characteristics of the system. However, its complete description is typically unknown, implying the value of $q$ as phenomenological \cite{SILVA2013201}. Thus, this generalization makes the model more versatile and capable of dealing with deviations from standard behaviors, since $q$ is capable of describing the degree of deformation of the exponential function. The results of the model are described in section \ref{sec:results}. 

\section{Variational Quantum Simulation of open systems}
\label{sec:3}

Quantum algorithms for simulating open quantum systems have been proposed in order to explore quantum properties in genuinely quantum logical architectures, according to Feynman's picture \cite{feynman2018simulating, lloyd1996universal}. Its simulations using quantum computing were described by Lloyd in the first concrete digital quantum simulation algorithm proposed \cite{lloyd1996universal}. The main problem in this approach is that, for simulating an environment effect, one needs to use more qubits or use the computing environment to take the same functional form as the coupling of the system to its environment, which requires more computational resources or a more controlled quantum architecture, respectively. 

An alternative to these problems is the construction of Variation Quantum Algorithms \cite{cerezo2021variational}, a class of hybrid quantum-classical algorithms with a more accurate simulation in the Noisy Intermediate-Scale Quantum (NISQ) era \cite{preskill2018quantum}. Recent work has proposed the use of a variational algorithm to simulate the dynamics of system dissipation due to the environment simply by using the CPU, called Variational Quantum Simulation (VQS) \cite{yuan2019theory}. Initially, the algorithm was proposed in \cite{li2017efficient} for closed systems simulation with active error minimization, but it was generalized in a series of applications in \cite{endo2020variational}, including open quantum systems simulation. 

Instead of simulating the evolution of the density matrix, VQS adopts an equivalent approach using the stochastic Schrödinger equation. This method models the evolution as an average over pure-state trajectories driven by continuous measurement processes. The use of the Stratonovich form of the linear Quantum State Diffusion (QSD) was proposed in \cite{luo2024variational}, leading to the state evolution. 
 
\begin{equation} \label{eq:Heff}
    \frac{\partial \ket{\psi_t}}{\partial t} = \left ( -iH - \frac{1}{2}\sum_k J_k L^{\dagger} L  + \sum_k z_k^*(t) L_k  \right ) \ket{\psi_k} 
\end{equation}
\begin{equation*}
    = -iH_{eff} \ket{\psi_t} \ .
\end{equation*}

The pure state evolution depends on a set of complex random variables \(z := \{z_k(t)\}\), which follow Gaussian white noise. These variables satisfy the statistical conditions \(\langle z_k(t) z_m^*(s) \rangle_z = \delta_{km} \delta(t - s)\) and \(\langle z_k(t) z_m(s) \rangle_z = 0\), where \(\langle \cdots \rangle_z\) denotes the average over the noise realizations \(z\). The reduced density matrix of the system, \(\rho(t)\), can be obtained by averaging over the stochastic trajectories of the pure state, as \(\rho(t) = \langle |\psi_t(z)\rangle \langle \psi_t(z)| \rangle_z\). Thus, the state evolves under the generalized time evolution with operator $H_{eff}$, which is equivalent to simulate the LME. Thus, VQS performs this operation based on the state evolution.
\begin{equation}
    \ket{\Psi (\Theta)} = \alpha \ket{\psi(\vec \theta)} = \alpha U(\vec \theta)\ket{\psi}_0  \ ,
\end{equation}
where $\Theta : = \{\alpha, \vec \theta \}$ represents the parameters used in the evolution state with $\alpha$ being the norm parameter that satisfies $\alpha^2 = \langle \Psi(\Theta) \ket{\Psi (\Theta)}$. Here, $U(\vec \theta)$ is the ansatz used in the circuit. For quantum simulation, a physically inspired ansatz is a natural choice since it allows mapping the system evolution, being able to avoid barren plateaus more efficiently if compared to a hardware-efficient ansatz \cite{park2024hamiltonian}. Thus, a useful ansatz class for VQS is the Hamiltonian variational ansatz (HVA), constructed as

\begin{equation}
    U(\vec{\theta}) = \prod_{j=1}^q e^{iH^{(j)} \theta_j^{l}} \ ,
\end{equation} 
with $l$ being the l-th layer and $H^{(j)}$ comes from the decomposition of the Hamiltonian into $m$ terms $H = \sum_{j=1}^m = c_jH^{(j)}$ with $\{c_j\} \in {\rm I\!R}$. 

Therefore, the system evolution can be mapped in terms of the parameter evolution using a variational principle. In \cite{yuan2019theory}, it was shown that McLachlan's principle \cite{mclachlan1964variational} can achieve better results for quantum evolution using VQA since it can be simply represented as

\begin{equation}
    \delta \left | \left | \frac{d}{dt} + iH_{eff} \right | \right | \ket{\Psi(\Theta)} = 0 \ .
\end{equation}

This is equivalent to computing the trajectory equation \cite{okuma2022nonnormal}

\begin{equation} \label{eq:trajetory}
    \sum_j^l M_{i, j} \dot \Theta_j = V_j \ ,
\end{equation}

where

\begin{align} \label{eq:parameters}
    M_{i,j} &= \operatorname{Re}\left(  \frac{\partial \bra{\Psi(\vec{\Theta}(t))}}{\partial \Theta_i} \frac{\partial \ket{\Psi(\vec{\Theta}(t))}}{\partial \Theta_j}  \right), \\
    V_j &= -\operatorname{Im}\left( \frac{\partial \bra{\Psi(\vec{\Theta}(t))}}{\partial \Theta_j} H_{eff} \ket{ \Psi(\vec{\Theta}(t))}    \right).
\end{align}

Both equations can be implemented in a quantum circuit using the Hadamard test \cite{yuan2019theory}, while Eq. \eqref{eq:trajetory} can be classically solved using an algorithm such as the fourth-order Runge-Kutta method. The techniques implemented imply an active error minimization, which allows it to simulate the evolution state with fewer layers than other methods, like the widely known Trotterization \cite{endo2020variational}. 

However, tests performed using VQS consider simpler interactions, without analyzing thermal interactions between the state and the environment \cite{yuan2019theory, endo2020variational, li2017efficient, luo2024variational}. In this work, we proposed the use of Eq. \eqref{eq:rho_open} in order to get a most complete description of $\hat \rho(t)$ for thermal states. The key idea is to use the norm parameter $\alpha$ to get a description of the probability decay $p(t)$. Since $\alpha^2 = \langle \Psi(\Theta) \ket{\Psi (\Theta)}$, we obtain $\alpha^4 \approx (1- e^{-\lambda t})$, satisfying $\alpha(0) = 1$ and $p(0)=0$. Therefore, since the quantum circuit only simulates the evolution of closed systems, the simulation was made using the parameters $\vec \theta$ to get the evolution $\{\rho_{00}(\vec \theta_t), \rho_{01}(\vec \theta_t), \rho_{10}(\vec \theta_t), \rho_{11}(\vec \theta_t)\}$, while the norm parameter $\alpha$ was use to compute the decay probability $p(t)$. Thus, Eq. \eqref{eq:rho_open} becomes

\begin{equation} \label{eq:rho_VQS}
    \hat{\rho}(t) =  \begin{pmatrix}
1 - \rho_{11} (\vec \theta_t) \cdot \alpha^4 + (1 - \alpha ^2) f & \rho_{01} (\vec \theta_t) \cdot \alpha^2 \\
\rho_{10} (\vec \theta_t) \cdot \alpha^2 & \rho_{11} (\vec \theta_t) \cdot \alpha^4 + (1- \alpha ^2) f \\
\end{pmatrix} \ .
\end{equation}

\begin{figure*}
    \centering
    \includegraphics[width=1\linewidth]{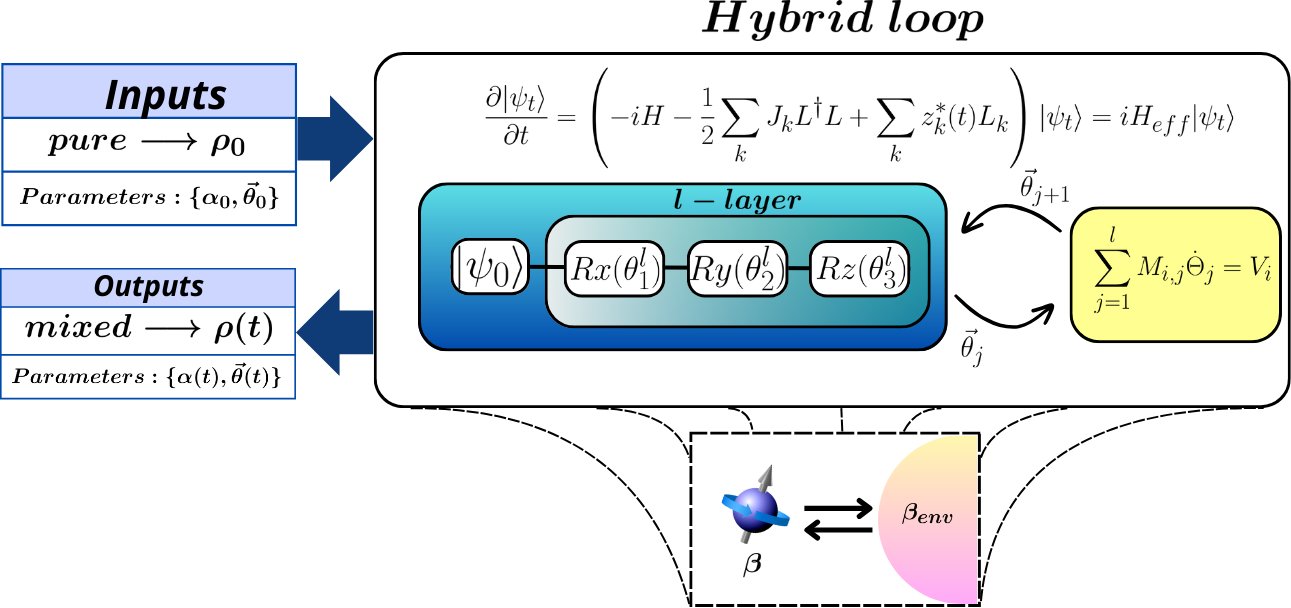}
    \caption{ \justifying VQS workflow. Initially, the pure closed system $\ket{\psi_0}$ and the initial parameters $\Theta_0$ are passed to the optimization process as inputs. At the end, the algorithm outputs are the optimized parameters $\Theta^*$ and state evolution $\ket{\psi(\vec \theta^*)}$, used to build the density matrix based on Eq. \eqref{eq:rho_open}.}
    \label{fig:VQS}
\end{figure*}

Therefore, this approach makes the algorithm capable of simulating $\rho(t)$ evolution using the physics of thermal states to build the density operator. The whole algorithm implementation workflow is shown in Figure \ref{fig:VQS}.

\section{Results and Discussion \label{sec:results}}

VQS performs quantum simulation considering the operator evolution expressed in Eq. \eqref{eq:Heff}. To simulate a thermal state interacting with a finite-temperature bath, the effective Hamiltonian is constructed from the dissipator in Eq. \eqref{eq:env}, written as

\begin{align} 
    H_{eff} = &\ H - \frac{i J}{2} \bigg\{ \frac{(1 - f)}{2} \left( \mathcal{I}  - \sigma_z \right) 
    + \frac{f}{2} \left( \mathcal{I} + \sigma_z \right) \notag \\ 
    &\ + z_1^* \sqrt{(1-f)} \sigma_- + z_2^* \sqrt{f} \sigma_+ \bigg\} \ ,
\end{align}
where $H$ is the Hamiltonian system. 

The characteristic noise of the QSD process was generated considering the techniques applied in \cite{luo2024variational} from the Ornstein-Uhlenbeck process \cite{gillespie1996exact},

\begin{equation}
    z(t + \Delta t)    = z(t) e^{-\gamma \Delta t} + \sqrt{\frac{\gamma}{2}(1-e^{-2\gamma \Delta t})}u \ ,
\end{equation}
where $u \sim \mathcal{CN}(0, 1)$ and $z(0) \sim \mathcal{CN}(0, \gamma/2)$. Thus, the expectation value of the operators is rewritten considering the averaging over $n_{traj}$ stochastic trajectories

\begin{equation} \label{eq:obs}
    \langle \hat{O} \rangle = \frac{1}{n_{\text{traj}}} \sum_{j=1}^{n_{\text{traj}}} \langle \psi_t(z) | \hat{O} | \psi_t(z) \rangle_j \ .
\end{equation}

Thus, we consider a total of $10^4$ trajectories with a time-step of $\Delta t = 0.15$. To ensure that the system starts as a pure state in a closed quantum system, we initialize it with $\alpha = 1$, since in the Bloch sphere representation a state with $|\alpha|^2 = 1$ corresponds to a pure state located on the surface of the sphere, whereas $|\alpha|^2 < 1$ characterizes a mixed state residing within the sphere. Furthermore, we choose the initial parameters $\vec \theta_0 = \{\theta_1^i = \theta_2^i = \theta_3^i = 0\}$ to be sent to the optimization process, so $\ket{\psi_0}$ is initially in the ground state. Therefore, the dissipation dynamics can be interpreted as an interaction between a spin-1/2 with $T(t =0) = 0 \ K$ and a hot bath for $f \neq 0$, which results in $T_{env} \neq 0$. 

In the following, we analyze the system's behavior under two distinct time-dependent Hamiltonians designed to probe the influence of thermal activation on qubit dynamics. Both configurations incorporate dissipative effects via a finite-temperature environment, but differ in the way the external driving field is modulated. The first Hamiltonian employs a conventional harmonic drive of the form \(\cos(\omega t)\), commonly used in qubit control protocols. The second introduces a composite generalized field that directly incorporates the activation dynamics through a nontrivial time dependence. These two driving schemes are explored separately in the following subsections to assess how the structure of the driving field interacts with different relaxation regimes and impacts the accuracy of the variational simulation.

\subsection{Oscillatory driving field}

A qubit can be effectively modeled as a fermionic system whose dynamics are influenced by the application of external magnetic fields~\cite{GeorgescuRevModPhys, christou2000single}. Thus, a more general control scheme for a qubit can be described by the Hamiltonian
\begin{equation} \label{eq:hamiltoniana}
    H(t) = g \mu_b B_{DC} \sigma_z + g \mu_b B_{AC} cos(\omega t) \sigma_x,
\end{equation}
where $B_{DC}$ is the DC field strength related to the Zeeman Hamiltonian, $B_{AC}$ is the maximum AC field strength applied in the x-axis, $g$ the isotropic Landré factor, $\mu_b$ represents the Bohr magneton. Here, $B_{DC}>B_{AC}$ ensures efficient control of the qubits. Research shows it is possible to build universal quantum gates applying magnetic fields, which allows the qubit to move around the entire Bloch's sphere for \cite{santos2017quantum, Oshima2003Proper}. 

Given this protocol, we set the parameters as $J = J_1 = J_2 =1$ for a moderate coupling with $B_{DC} = 2 \ \text{mT}$ and $B_{AC} = 0.5 \ \text{mT}$, and $T_{\text{env}} = 10 \ \text{K}$ to capture quantum regimes. First, we evaluate the populations of the quantum system under the conditions described in regime \ref{item:i} and \ref{item:ii} to analyze different limits with fixed frequencies. We perform it in a noisy quantum simulator described in Appendix \ref{app:noise} to compute the expected values following the definition in Eq. \eqref{eq:obs}, in order to visualize the algorithm's efficiency in a more realistic scenario. For the quantum simulation, we use a 3-layer ansatz shown in Fig. \ref{fig:VQS}, while the exact solutions were computed using \textit{qutip} --- an open-source Python framework for the dynamics of open quantum systems \cite{johansson2012qutip}. 

\begin{figure*}[ht!]
 \centering
  \subfloat[ ]{
    \includegraphics[width=.5\linewidth]{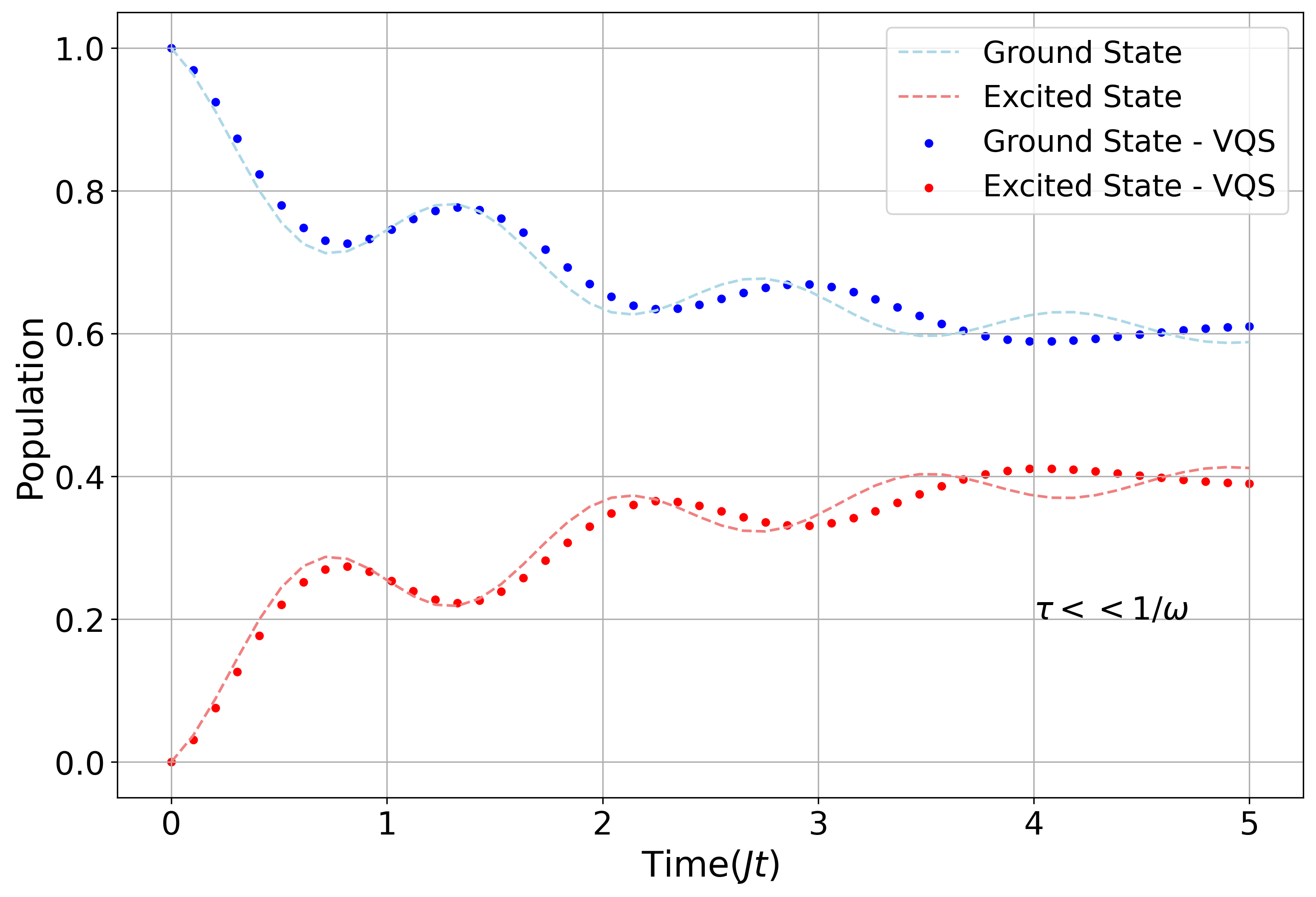}
  }
  \subfloat[ ]{
    \includegraphics[width=.5\linewidth]{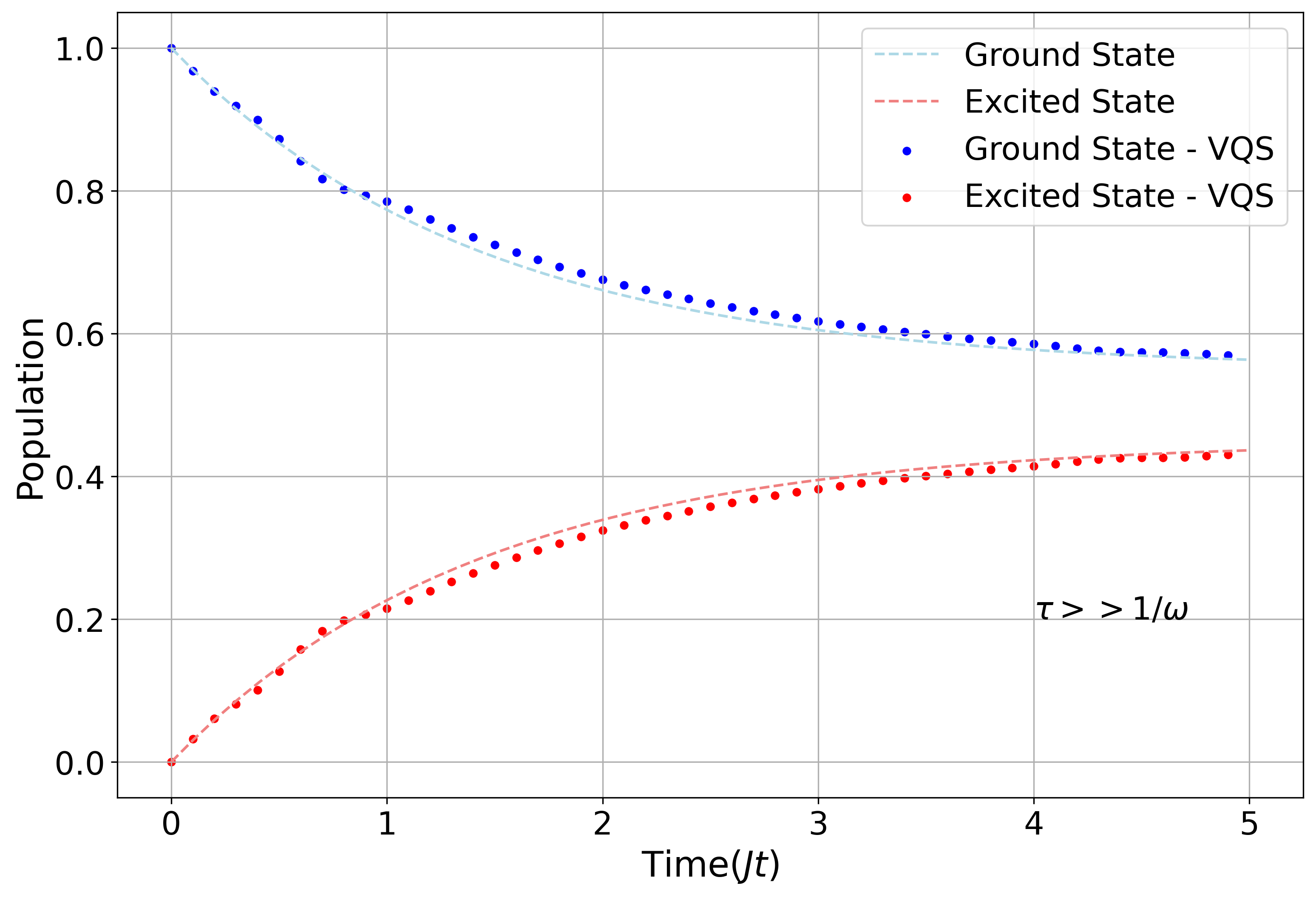}
  } \\
  \subfloat[ ]{
    \includegraphics[width=.5\linewidth]{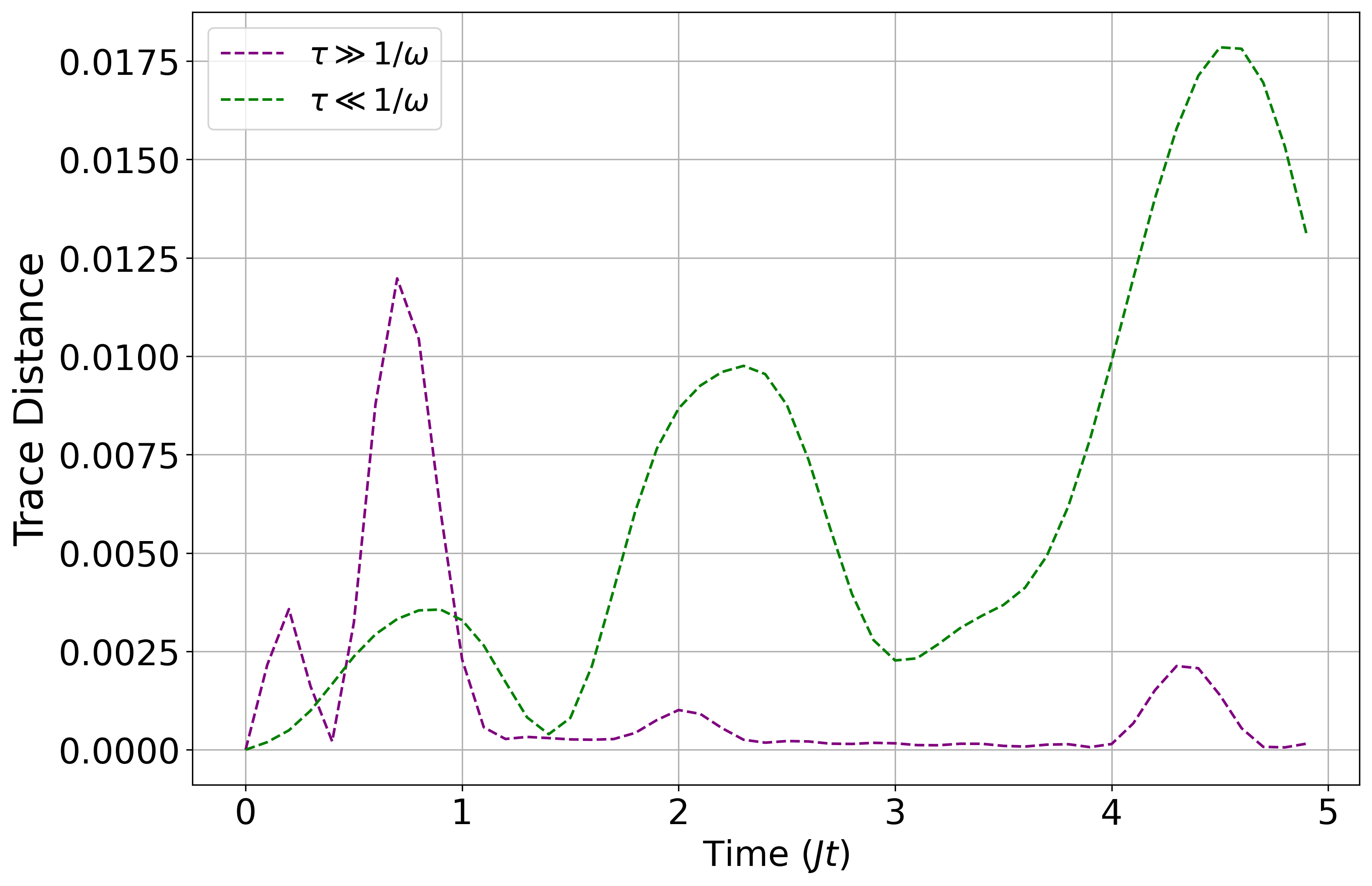}
  }
  \caption{\justifying Regime \ref{item:i} and \ref{item:ii} using VQS for quantum simulation (dots) and qutip for exact results (dashed). (a) Simulation for regime \ref{item:i} with $\tau \gg 1/\omega$. (b) Simulation for regime \ref{item:ii} with $\tau \ll 1/\omega$. (c) Trace distance between the simulated (dots) and exact (dashed) results for both regimes.}
  \label{fig:regime_i_ii}
\end{figure*}

The results of the simulation for both regimes can be visualized in Fig. \ref{fig:regime_i_ii}, with the points representing the average in each time step. Additionally, there is a trace distance between the exact and simulated density operator, reflecting the noise effect and the algorithm's precision. In the regime where $\tau \ll 1/\omega$, the system displays quick oscillations with underdamping, while it exhibits overdamping behavior for $\tau \gg 1/\omega$. The exact results agree with the behavior described before from the reference \cite{topping2018ac}, and the simulation also shows agreement with the exact result: for $1/\tau \gg \omega$, the algorithm is able to capture the general behavior of the system, except for small oscillations; for $1/\tau  \gg \omega$, the system shows a more effective correspondence. 

\begin{figure*}
     \centering
     \includegraphics[width=1\linewidth]{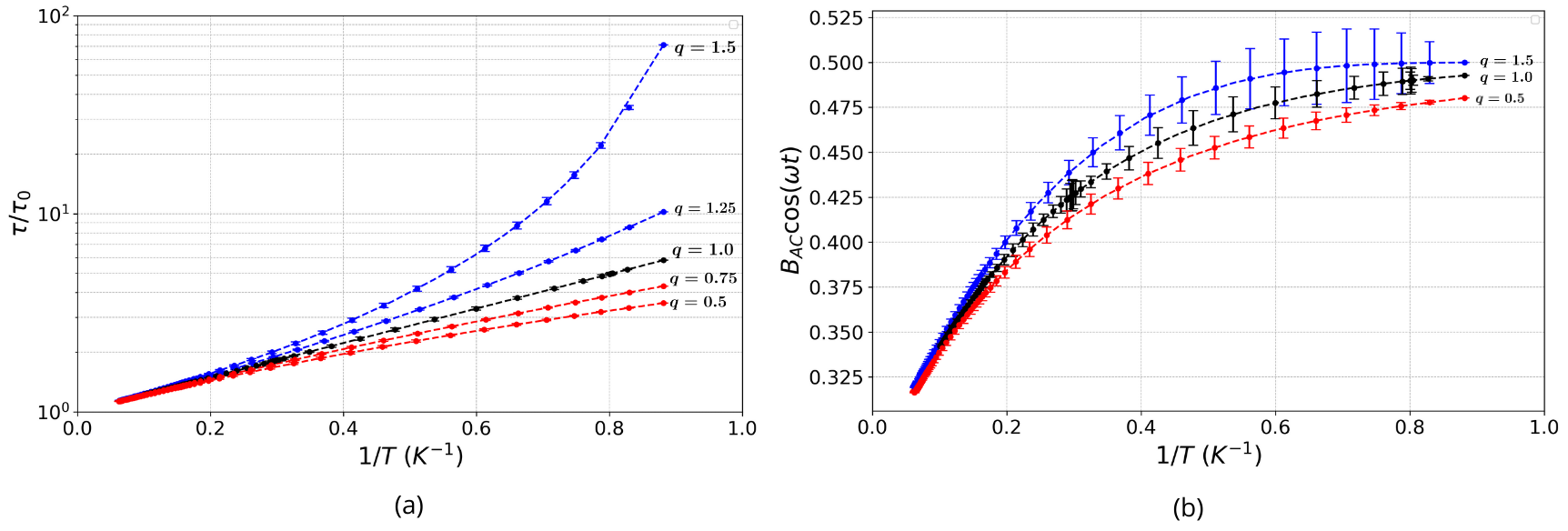}
      \caption{ \justifying  Nonadditive model using VQS for noisy quantum simulation (dots with error-bar) and qutip for exact results (dashed). (a) Relaxation time considering Eq. \eqref{eq:nonad} for $q=\{0.5, 0.75, 1, 1.25, 1.5\}$. (b) The magnetic field AC influence on the system Hamiltonian in Eq. \eqref{eq:hamiltoniana}, considering the relaxation time. The results show that the nonadditive dynamics have little influence on the Hamiltonian of the system.}
     \label{fig:time}
 \end{figure*}

The efficacy of the simulation makes it possible to study other models with more complex dissipation effects, considering the nonactive model proposed in Eq. \eqref{eq:nonad}. For this purpose, the regime \ref{item:iii} is more appropriate to study non-Arrhenius models, since we can be directly related to the field frequency $\omega$. Thus, we simulate the model considering the configurations described before for $q = \{0.5, 0.75, 1, 1.25, 1.5\}$. The main result from the nonadditive model with the oscillatory driving field can be visualized in Fig. \ref{fig:time}, with the relaxation time for the different chosen parameters and the influence of their different dynamics on the AC magnetic field.

The first result from the nonadditive model simulation is the existence of two different regimes. In the nonadditive model proposed in \cite{rosa2019characterization}, the authors describe the non-Arrhenius behaviors as concave curves (sub-Arrhenius behavior), linked to non-local quantum effects \cite{Antonio2011, jp503463w, SILVA2013201}, or as convex curves (super-Arrhenius behavior), associated with the dominance of classical transport phenomena \cite{PhysRevBChamberlin, NISHIYAMA2009325, Roy2017, Carvalho24529}. These results show a correspondence with the phenomenological relationships proposed to model non-Arrhenius processes, such as the Vogel-Tamman-Fulcher equation \cite{demsar2020non, erdem2011nonequilibrium, blackmore2023characterisation} and the Aquilanti-Mundim d-Arrhenius model \cite{NISHIYAMA2009325, Roy2017, Carvalho24529}. Figure \ref{fig:time} shows that the curves have the same results in terms of characterization of sub or super-Arrhenius behavior but different associations to relaxation time phenomena. 

Henceforth, we will discuss possible interfaces between the nonadditive model proposed in Eq. \eqref{eq:nonad} and existing models in the literature for relaxation time with thermal activation. However, before introducing the different relaxation regimes used in this work, we emphasize that their detailed physical interpretation is beyond the scope of this study. Our primary goal is to investigate how such generalized dissipative behaviors affect the performance and adaptability of variational quantum simulations. Thus, the key idea is to introduce the phenomenological relationships that do not appear in the Arrhenius-type model in order to observe its influence in quantum simulation. 

In the sub-Arrhenius regime ($q>1$), the system is slightly influenced by temperature, having a slower decrease. This behavior was mapped in different models, such as Heitler-Teller in the study of metals subjected to the surface of the Fermi distribution \cite{heitler1936time}. The authors showed that metals with high magnetic moments also have a short relaxation time at low temperatures. This behavior was also observed in other studies \cite{korringa1950nuclear, erdem2011nonequilibrium}. Although these characteristics are reported in specific models, they appear naturally in the nonadditive model, which shows a certain phenomenological correspondence.

In the super-Arrhenius regime ($q<1$), an important fact is the existence of transition temperatures in the model. Different models describe the critical temperatures present in relaxation time, widely used in systems with phase transitions, such as the Vogel-Tamman-Fulcher equation \cite{demsar2020non, erdem2011nonequilibrium}, Basser’s law \cite{blackmore2023characterisation}, or the so-called Critical Slowing Down model \cite{yahata1969critical, fisher1986scaling}. These models represent characteristics already observed in certain materials, such as the dynamics of an interacting particle system \cite{djurberg1997dynamics}, the existence of a complex magnetic ground state \cite{yadav2023existence}, and abrupt transition behavior \cite{kundu2020critical}. Again, these characteristics appear naturally in the nonadditive model, a behavior not observable in the Arrhenius-type model.

In this sense, exploring systems governed by nonadditive relaxation dynamics within the VQS framework offers a valuable opportunity to assess the algorithm’s capability to generalize across different dissipative regimes. Since many physical systems exhibit deviations from standard Arrhenius-like behavior, it becomes essential to understand how such variations in the relaxation profile impact the performance and expressibility of variational algorithms. By simulating dynamics that go beyond the exponential decay typical of additive models, we can evaluate whether the variational ansatz remains robust under structurally distinct forms of dissipation. This approach not only helps benchmark the versatility of VQS in modeling open quantum systems but also sheds light on the types of relaxation dynamics that are more naturally compatible with variational representations.

\begin{figure*}[ht!]
 \centering
  \subfloat[ ]{
    \includegraphics[width=.5\linewidth]{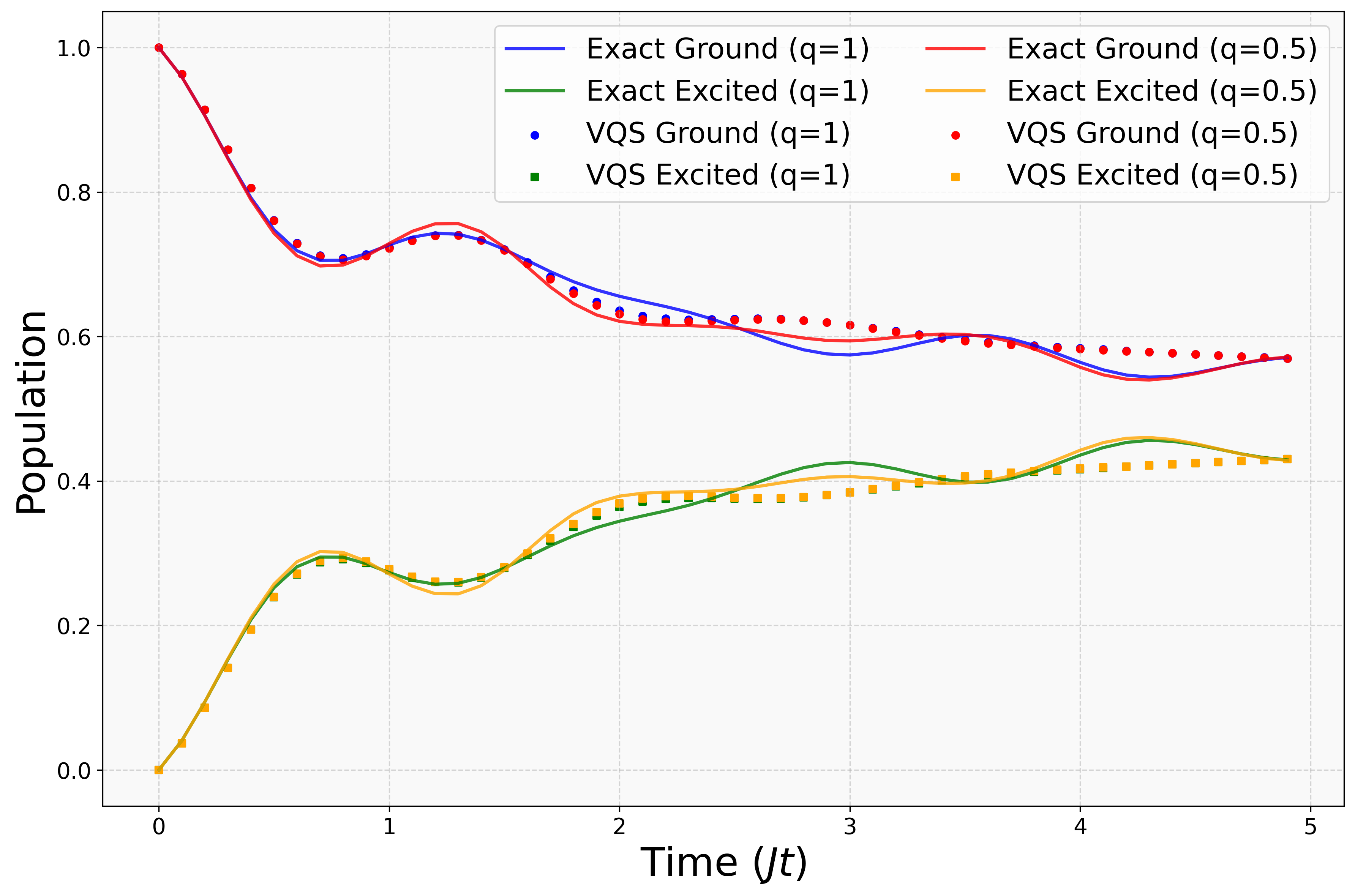}
  }
  \subfloat[ ]{
    \includegraphics[width=.5\linewidth]{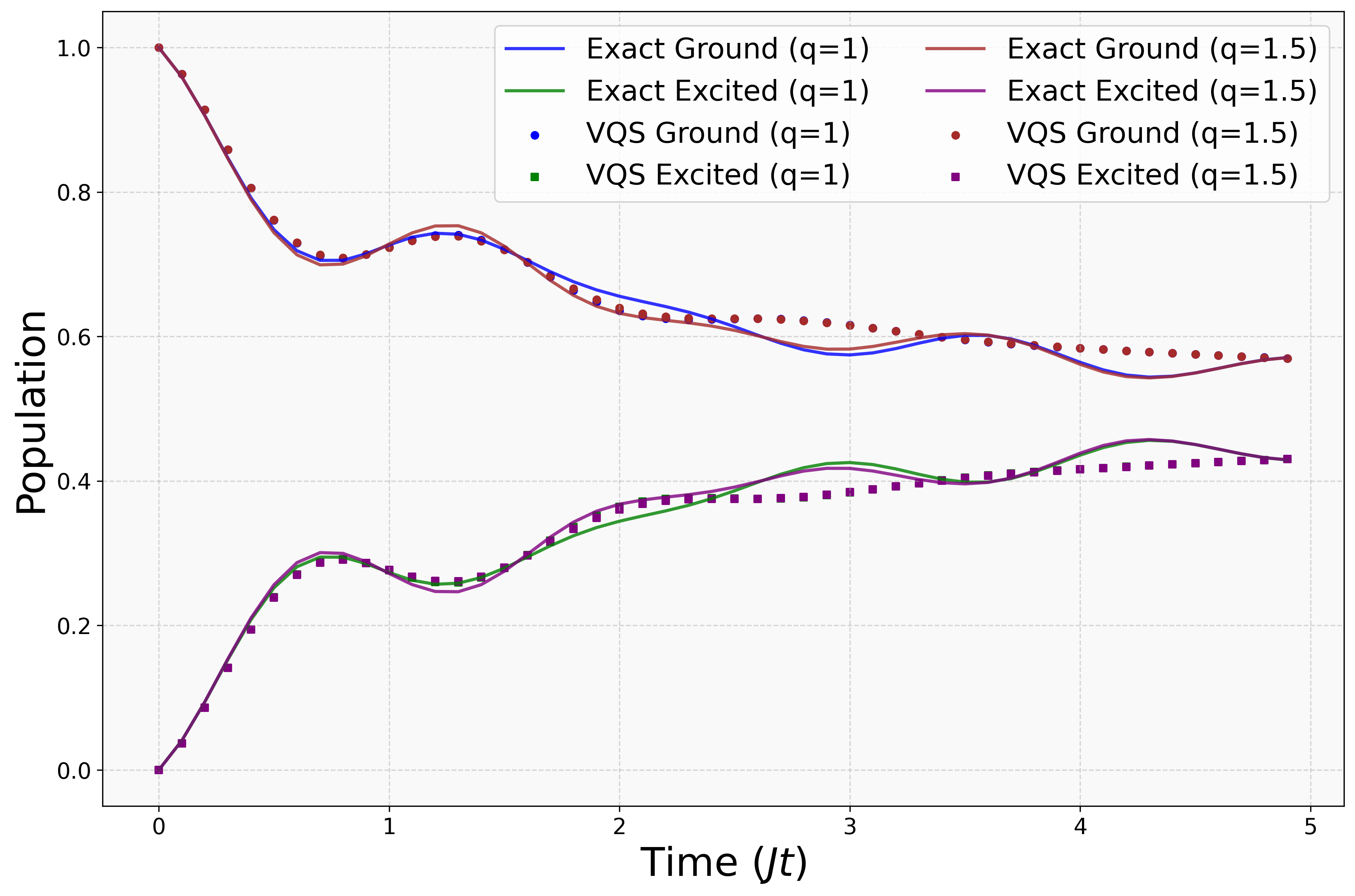}
  } \\
  \subfloat[ ]{
    \includegraphics[width=.5\linewidth]{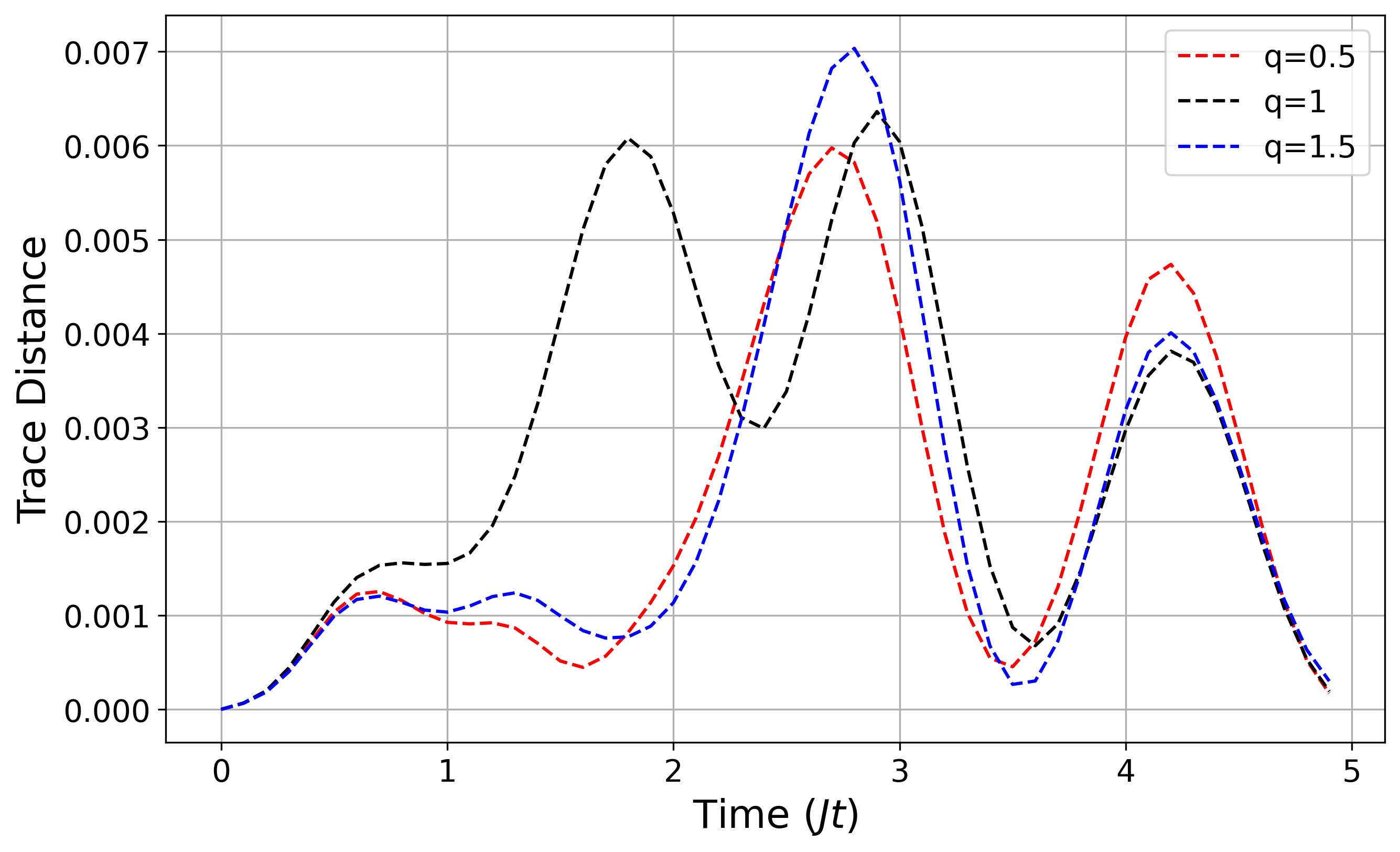}
  }
  \caption{\justifying Populations of $\rho(t)$ considering the nonadditive model in regime \ref{item:iii} simulated using VQS (dots) and qutip for exact results (dashed lines) considering the oscillatory driving field described in the Hamiltonian presented in Eq. \eqref{eq:hamiltoniana}. a) Population of $\hat \rho(t)$ considering $q=1$ (Arrhenius-type) and $q=0.5$ (super-Arrhenius). b) Population of $\hat \rho(t)$ considering $q=1$ (Arrhenius-type) and $q=1.5$ (super-Arrhenius). c) Trace distance between the quantum simulation using VQS (dots) and exact results using qutip (dashed lines).}
  \label{fig:regime_iii}
\end{figure*}

In the simulation, Fig. \ref{fig:time} also has an important result in terms of the influence of the non-Arrhenius regime on the external AC field for different temperatures. Usually, the Zeeman part of the Hamiltonian $H$ in Eq. \eqref{eq:hamiltoniana} has a predominant effect in the evolution of a qubit since $B_{DC} \geq {B_{AC}}$, besides the fact that $B_{AC} \cos{(\omega t)} \in [-B_{AC}, B_{AC}]$. This characteristic results in the fact that, even though the relaxation time is affected by different dynamics, the evolution of the density matrix does not suffer significant influence under the regime \ref{item:iii}. In order to visualize this behavior, we plot the population from $\rho(t)$ for different parameters in Fig. \ref{fig:regime_iii}.

The results show a curve with an intermediary behavior from regime \ref{item:i} and \ref{item:ii}. As expected, different values of $q$ do not lead to significant changes, except for the deformation of the curves: the exact results show only small deviations, while the simulated ones remain nearly unchanged. However, Fig. \ref{fig:regime_iii} also shows that the trace distance is reduced for certain values of $q$. Thus, the simulation reveals that VQS can mimic behaviors of non-additive systems with greater fidelity in specific cases by generalizing the system's relaxation time.

\subsection{Composite generalized field}

The oscillatory driving field does not respond significantly to the generalization of the time parameter $1/\tau_q$. Thus, we use another controlled mechanism for the AC magnetic field with direct influence of the frequency capable of providing a valuable reference baseline for understanding the system's sensitivity, which can be written as

\begin{equation} \label{eq:hamiltoniana2}
    H(t) = g \mu_b B_{DC} \sigma_z + g \mu_b h(t) \sigma_x \ .
\end{equation}

Here, the external field $h(t)$ is varied back and forth between two extreme values $ h_1$ and $ h_2$, considering the frequency of the system, while the system is kept in contact with the heat bath. Considering the regime \ref{item:iii}, one can simulate a more complex regime by applying a particular protocol of the form

\begin{equation}
    h(t) = \sqrt{h_2^2 (\omega t) + h_1^2(1-\omega t)} \ .
\end{equation}

\begin{figure*}
    \centering
    \includegraphics[width=1\linewidth]{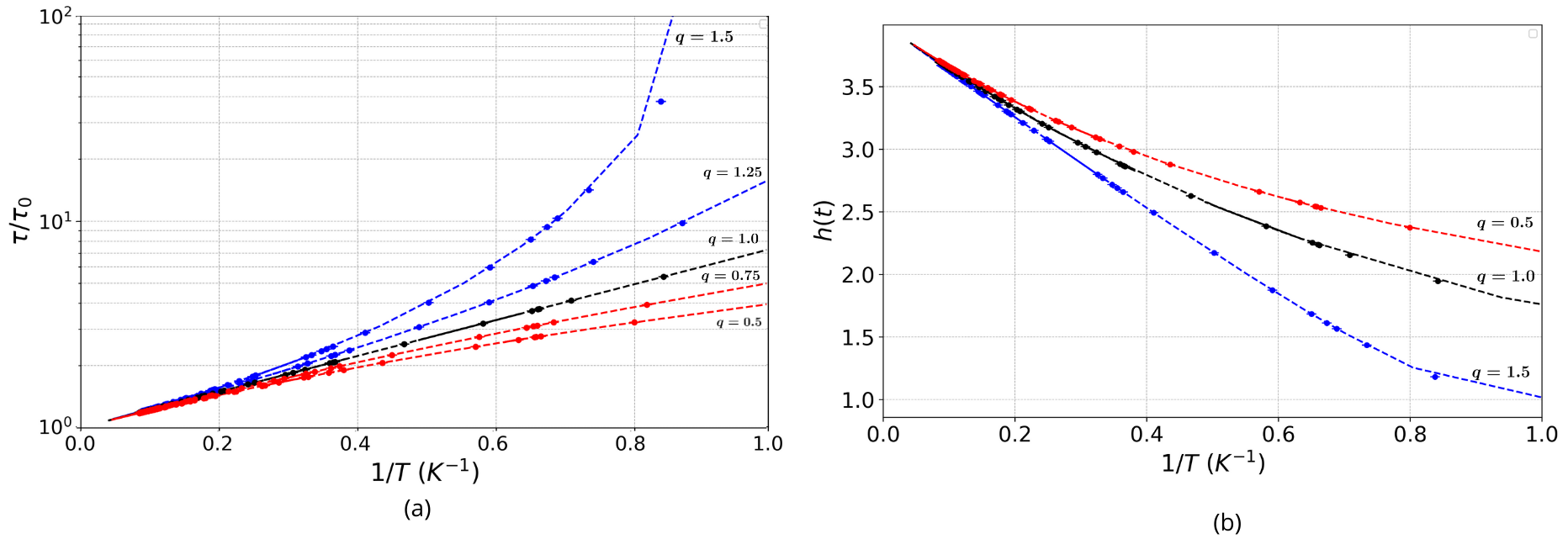}
  \caption{ \justifying  Nonadditive model using VQS for noisy quantum simulation (dots with error-bar) and qutip for exact results (dashed) considering the composite generalized field described in the Hamiltonian presented in Eq. \eqref{eq:hamiltoniana2}. Relaxation time considering Eq. \eqref{eq:nonad} for $q=\{0.5, 0.75, 1, 1.25, 1.5\}$ (Left). The magnetic field AC influence in the system Hamiltonian in Eq. \eqref{eq:hamiltoniana}, considering the relaxation time (Right). The results show that the nonadditive dynamics have little influence on the Hamiltonian of the system.}
  \label{fig:htime}
\end{figure*}

Given this definition, we set the parameters as $J = J_1 = J_2 =1$ for a moderate coupling with $B_{DC} = 4 \ \text{mT}$, $h_1 = 1$ and $h_2 = 3$, respecting $B_{DC}\geq h(t)$, and keep $T_{\text{env}} = 10 \ \text{K}$. Thus, we reproduce the same results now using the Hamiltonian in Eq. \eqref{eq:hamiltoniana2} in the effective Hamiltonian in Eq. \eqref{eq:Heff}. Figure \ref{fig:htime} shows the result for the relaxation time and $h(t)$ under a nonadditive dynamics. Here, the relaxation time displays the same behaviors observed using the oscillatory driving field, with the same regimes of sub- and super-Arrhenius described before. By contrast, this field reveals a strong dependence on $1/\tau_q$, changing its degree of decay based on the value of $q$. 

\begin{figure*}[ht!]
 \centering
  \subfloat[ ]{
    \includegraphics[width=.5\linewidth]{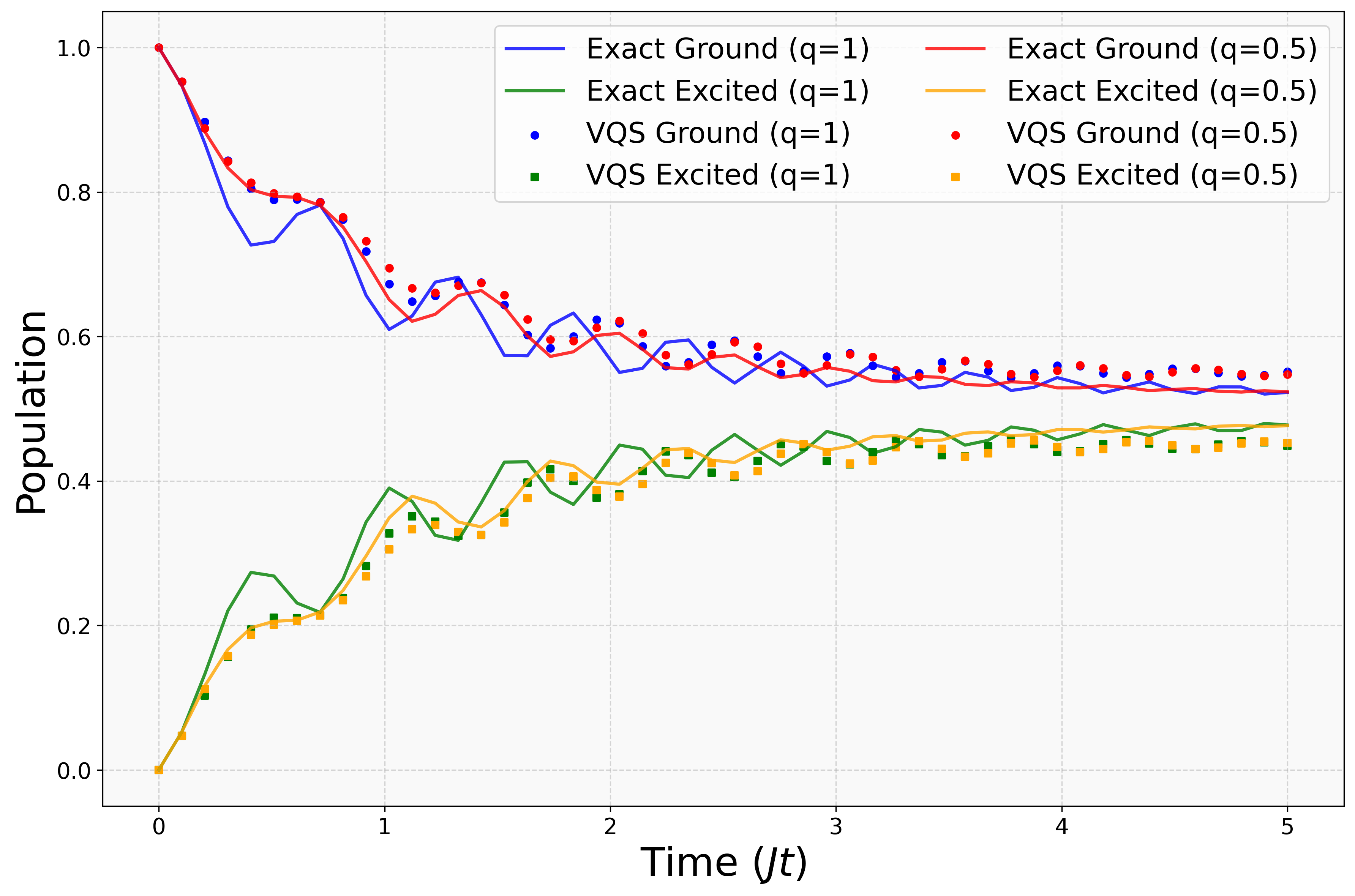}
  }
  \subfloat[ ]{
    \includegraphics[width=.5\linewidth]{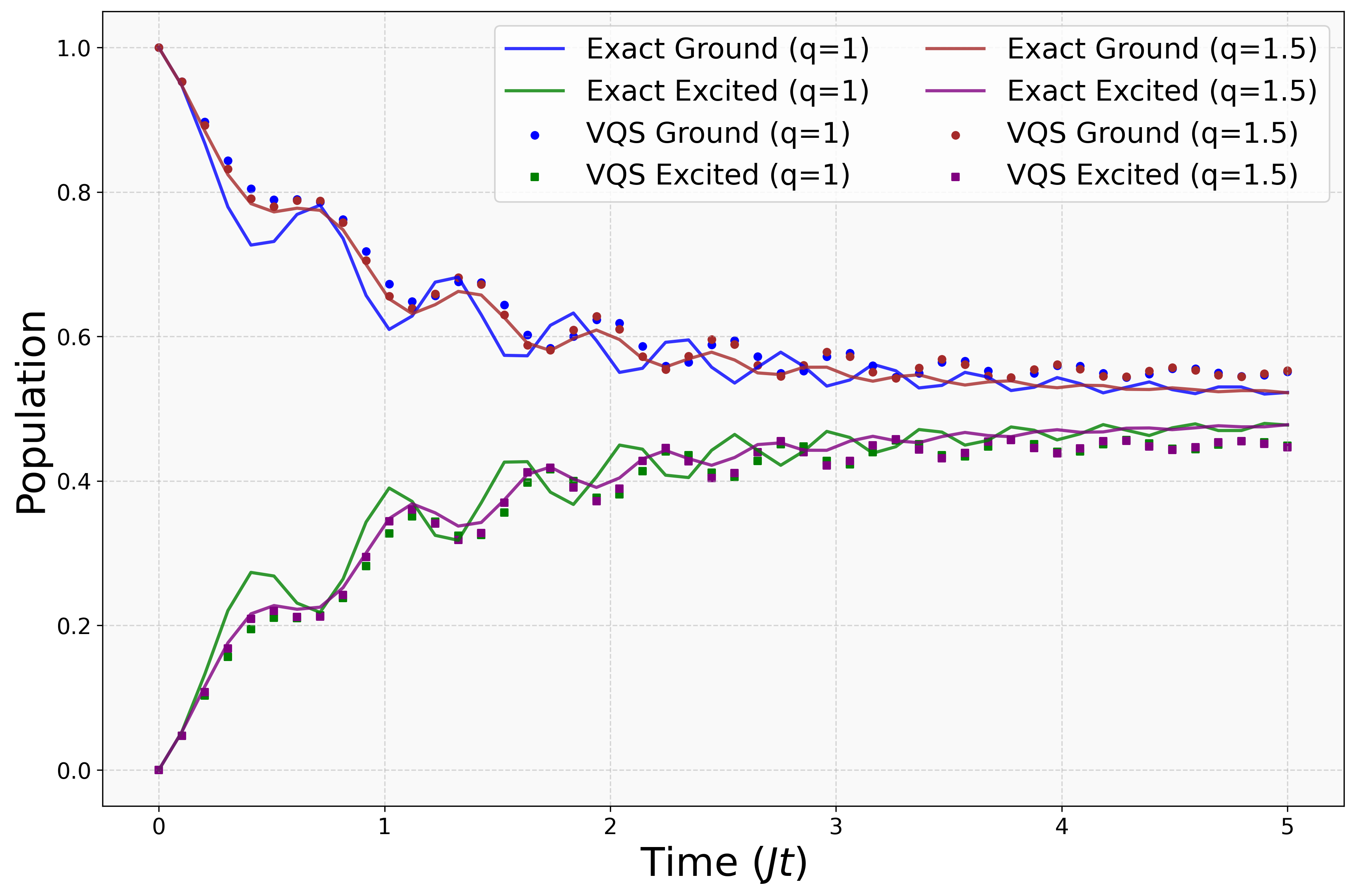}
  } \\
  \subfloat[ ]{
    \includegraphics[width=.5\linewidth]{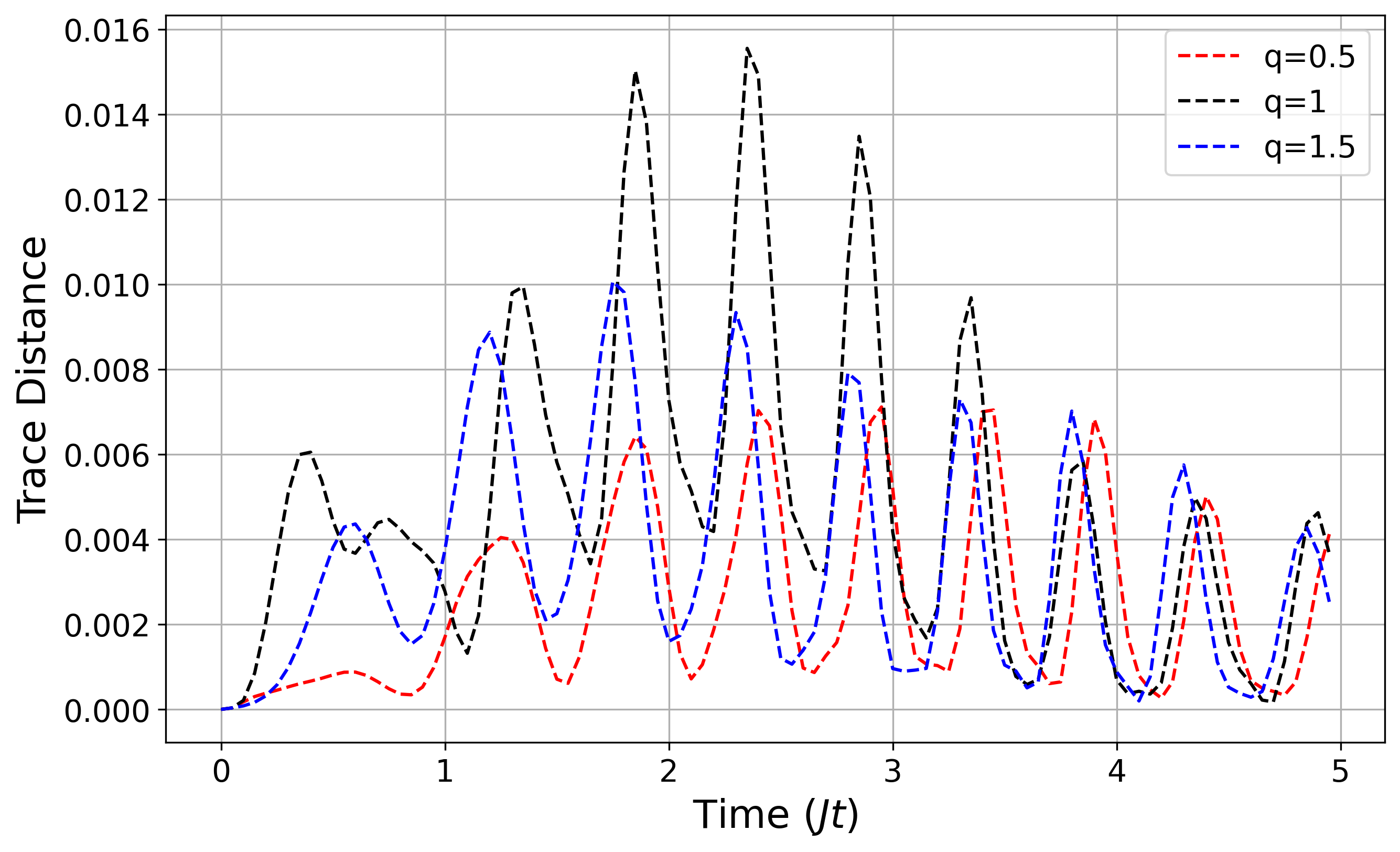}
  }
  \caption{\justifying Populations of $\rho(t)$ using the nonadditive model in regime \ref{item:iii} simulated using VQS (dots) and qutip for exact results (dashed lines) considering the composite generalized field described in the Hamiltonian presented in Eq. \eqref{eq:hamiltoniana2}. a) Population of $\hat \rho(t)$ considering $q=1$ (Arrhenius-type) and $q=0.5$ (super-Arrhenius). b) Population of $\hat \rho(t)$ considering $q=1$ (Arrhenius-type) and $q=1.5$ (super-Arrhenius). c) Trace distance between the quantum simulation using VQS (dots) and exact results using qutip (dashed lines).}
  \label{fig:pop-h}
\end{figure*}

These results make it possible to simulate the response of the system under different regimes. Figure \ref{fig:pop-h} shows the result for the populations, with visible differences. Varying the nonadditivity parameter $q$ modifies the envelope of the composite field $h(t)$, smoothing the transition step between its two branches. This smoother drive also tempers the instantaneous coupling to the thermal bath under the generalized amplitude-damping channel, yielding gentler, more uniform relaxation trajectories. Consequently, the VQS variational manifold --- which is optimized to capture the dominant, low-frequency features of the evolution \cite{yuan2019theory, endo2020variational, li2017efficient, luo2024variational} --- achieves higher mapping accuracy in non-Arrhenius regimes, where the narrower bandwidth of $h(t)$ aligns better with the expressibility of the ansatz.

The results indicate that the VQS algorithm can serve as an approximate mapping framework for exploring thermal activation dynamics, particularly in regimes that deviate from classical Arrhenius behavior. While the exact solution exhibits a pronounced dependence on the form of the relaxation law—particularly under nonadditive generalizations—the VQS output remains comparatively stable across different values of the generalization parameter $q$. This apparent insensitivity does not reflect a lack of resolution; rather, it suggests that the variational ansatz is capable of capturing the dominant features of the dynamics even when relaxation deviates from the exponential form. The reduced discrepancy between variational and exact results in non-Arrhenius regimes points to a structural compatibility between the variational manifold and the underlying effective dynamics. These findings reinforce the potential of VQS not only as a simulation tool but also as a probe for characterizing the qualitative nature of dissipative processes in open quantum systems.

Another important observation about the method proposed in Eq. \eqref{eq:rho_VQS} is that the definition can be used to estimate the coherence terms in $\hat \rho(t)$, which has not been explored in other works \cite{yuan2019theory, endo2020variational, li2017efficient, luo2024variational}. These terms are important to compute the coherence, given by $\mathcal{C}(t) = \sum_{i \neq j} |\rho_{ij}(t)|$. A more general picture of coherence along Bloch's sphere can be visualized using a basis transformation matrix $\mathcal{\hat U}$ to describe a state in a different basis, such that
\begin{equation} \label{eq:gencoherence}
    \mathcal{C}(\theta, \phi, t) = \sum_{i\neq j} |\rho_{ij}(\theta, \phi, t)| \ ,
\end{equation}
where $\rho(\theta, \phi) = \mathcal{\hat U} (\theta, \phi, t) \rho(t) \mathcal{\hat U}^{\dagger} (\theta, \phi)$ and 
\begin{equation}
    \mathcal{\hat U}(\theta, \phi) = \begin{pmatrix}
cos(\theta/2) &  -e^{i\phi}sin(\theta/2)\\
e^{-i\phi}sin(\theta/2) &  cos(\theta/2) \\
\end{pmatrix} \ .
\end{equation}
Equation \eqref{eq:gencoherence} makes it possible to compute the coherence in a single point of the Bloch's sphere. Consequently, it is also possible to compute the average coherence considering an integration along the whole sphere as

\begin{equation}
   \langle \mathcal{C}(t)\rangle = \frac{1}{4 \pi}\int_0^{2 \pi} \int_0^\pi \mathcal{C}(\theta, \phi, t)sin(\theta) d\theta d\phi \ . 
\end{equation}

\begin{figure}
    \centering
    \includegraphics[width=1\linewidth]{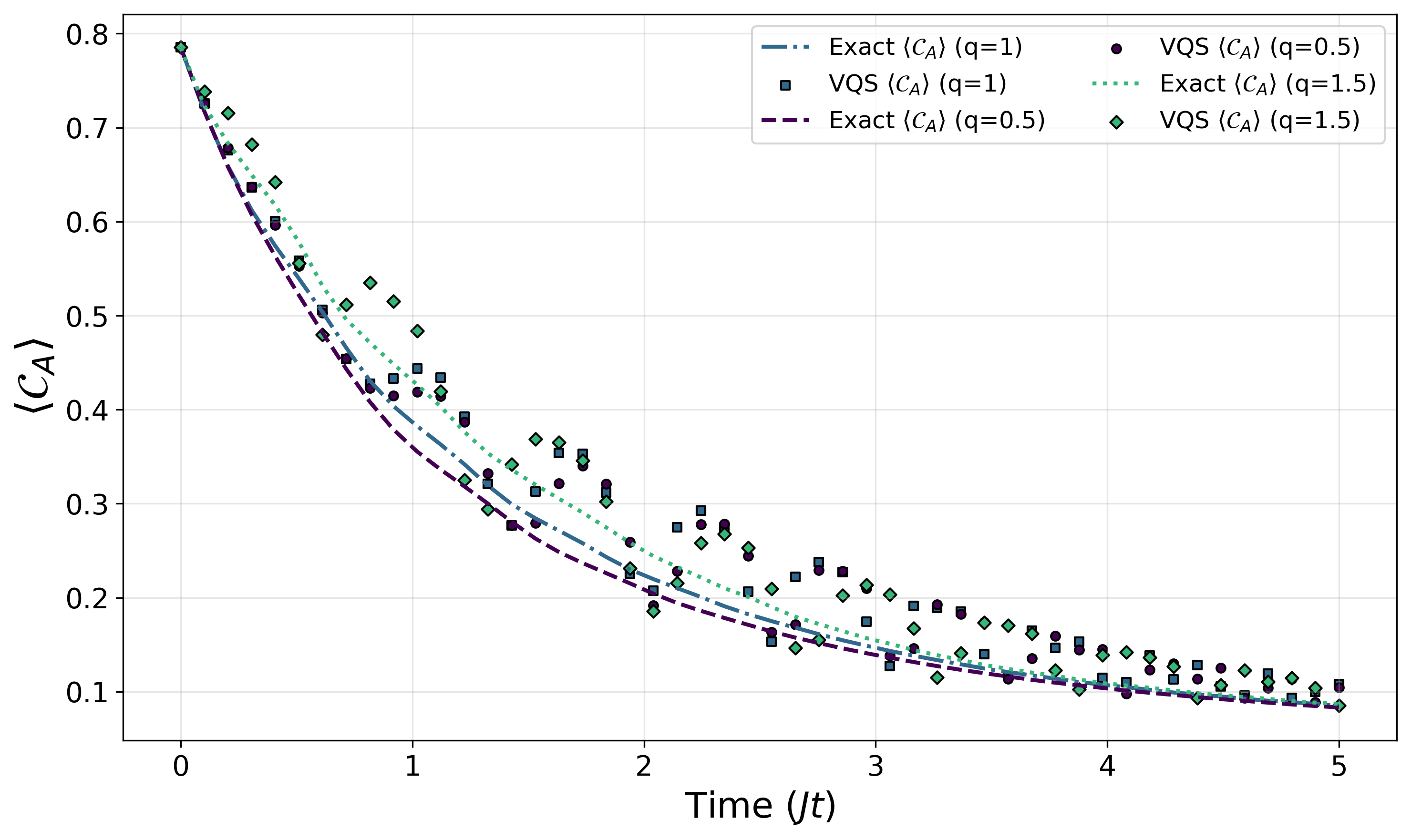}
    \caption{\justifying Average coherence in Eq. \eqref{eq:gencoherence} considering the nonadditive model in regime \ref{item:iii} simulated using VQS (dots) and qutip for exact results (dashed lines). As expected, the results show a gradual decay. }
    \label{fig:avcohe}
\end{figure}

Aiming to visualize the general correspondence, we plot the average coherence in Fig. \ref{fig:avcohe} considering the nonadditive model in regime \ref{item:iii}. The results show consistency with the expected, since the state lost coherence gradually: a quantum state in contact with an environment of temperature $10 \ K$ will result in a state with norm $\alpha = 0$, located at the center of Bloch's sphere. The different parameters $q$ imply different degrees of decay. The quantum simulation using VQS shows a slow decay, too, but with small oscillations. 

Therefore, the observed coherence decay supports the consistency of the nonadditive model in capturing the expected thermalization behavior of an open quantum system. The close agreement between the exact solution and the VQS simulation, despite minor oscillations in the variational trajectory, demonstrates the algorithm's capacity to reproduce both the qualitative and quantitative aspects of coherence loss under generalized dissipative conditions. These results further reinforce the utility of VQS as a viable approach for simulating the decoherence dynamics of open systems, particularly in scenarios where standard relaxation laws are extended to include more complex, nonadditive behavior.

\section{Conclusion}

\label{sec:5}

We employ a variational quantum simulation algorithm to analyze different regimes of a qubit subjected to a finite-temperature bath. As a first result, we simulated the Arrhenius law in Eq. \eqref{eq:arrh} under regimes \ref{item:i} and \ref{item:ii} to analyze the robustness of the algorithm considering noise. Additionally, we have proposed an alternative way to describe the relaxation time in a non-Arrhenius regime with thermal activation. Our model written in Eq. \eqref{eq:nonad} captures behaviors observed in existing models by introducing the phenomenological parameter $q$ in the Arrhenius-type thermal activation with approximately temperature-independent activation energy, demonstrating its potential towards a general description of relaxation time. This result was used to simulate the quantum dynamics of dissipative spin-1/2 under nonadditive regime. 

In summary, our work demonstrates that embedding a realistic, finite‑temperature dissipative model into the VQS framework --- and further extending it with a tunable, nonadditive relaxation law --- yields a powerful and flexible tool for characterizing open‑system dynamics on near‑term quantum hardware. By showing that the variational ansatz can map both standard and non‑Arrhenius activation regimes, and that smoother, generalized driving fields lead to reduced simulation error, we provide a blueprint for benchmarking and optimizing quantum simulators under physically relevant conditions. These results pave the way for more accurate studies of thermalization, decoherence, and energy‑exchange processes in quantum materials and devices, and highlight the potential of variational algorithms not only as computational engines but also as diagnostic probes of complex dissipative phenomena.


\begin{acknowledgments}
The authors thank Fundação de Amparo à Pesquisa do Estado da Bahia - FAPESB for its financial support (grant numbers APP0041/2023 and PPP0006/2024). LQG thanks QuIIN—EMBRAPII CIMATEC Competence Center in Quantum Technologies, with financial resources from the PPI IoT/Manufatura 4.0 of the MCTI grant number 053/2023, signed with EMBRAPII, for its financial support. Also, this work received partial financial support from CNPq (Grant Numbers: $305096/2022-2$ MAM). \\

\end{acknowledgments}

\appendix

\section{Applied Noisy Model}\label{app:noise}
The noise framework used in this work is built on a well-defined set of native operations, including \textit{cx}, \textit{delay}, \textit{id}, \textit{measure}, \textit{reset}, \textit{rz}, \textit{sx}, and \textit{x}, in which the quantum circuits are constructed, ensuring compatibility with the noise model. Among them, only part of this gate library is modeled as error-susceptible: namely \textit{sx}, \textit{id}, \textit{x}, \textit{cx}, and \textit{measure}. Attaching noise to this subset can capture the dominant physical processes for common decoherence mechanisms, such as pure dephasing, relaxation, and imperfections during the gate executions.

The specification targets two qubits, indexed as $0$ and $1$. Thus, all noise channels are applied to these particular registers rather than to an arbitrary qubit set. Each noisy operation can be associated with a specific single qubit (\textit{id}, \textit{sx}, \textit{x}, \textit{measure}) or a qubit pair (\textit{cx} gate). In particular, it is worth highlighting that including errors in the \textit{id} gate is especially important, since it mimics decoherence experienced by qubits that remain idle, primarily due to finite $T_1$ and $T_2$ relaxation times. Here, $T_1$ represents the energy-decay time, $T_2$ describes the phase-coherence time, where they are fixed as $T_1 = 0.00015774397097652505 \ s$ and $T_2 = 0.00010861203881817735 \ s$. The qubits operate near $5.23$ GHz,  a frequency typical of superconducting quantum computing platforms. Moreover, independent read-out error models are also assigned to the qubits, in order to reflect thermal noise and imperfections in the measurement electronic chain and the quantum-to-classical conversion processes.

In the following, Table \ref{tab:noise} presents the main characteristics of the used noise model, detailing the basis gates, the error-treated instructions, affected qubits, and other specific error sources. By integrating these parameters, the used model provides a wide representation of noise effects, offering a realistic scenario for the simulations under decoherence and gate inaccuracy.

\begin{table}[h]
\centering
\renewcommand{\arraystretch}{2} 
\setlength{\tabcolsep}{5pt} 
\begin{tabular}{@{}p{2.5cm}p{5.5cm}@{}}
\toprule
\textbf{Attribute}              & \textbf{Configuration}                                                                                   \\ \midrule
{Basis gates}            & \texttt{['cx', 'delay', 'id', 'measure', 'reset', 'rz', 'sx', 'x']}                                     \\ 
{Instructions with noise} & \texttt{['sx', 'id', 'x', 'cx', 'measure']}                                                             \\ 
{Qubits with noise}      & \texttt{[0, 1]}                                                                                         \\ 

{$T_1 (s)$}      & \texttt{0.00015774397097652505}                                                                                         \\ 

{$T_2(s)$}      & \texttt{0.00010861203881817735}  \\

{Frequency(Hz)}      & \texttt{5227644738.696302}  \\

{Specific qubit errors}  & \texttt{[('cx', (0, 1)), ('id', (0,)), ('id', (1,)), ('sx', (0,)), ('sx', (1,)), ('x', (0,)), ('x', (1,)), ('measure', (0,)), ('measure', (1,))]} \\ \bottomrule
\end{tabular}
\caption{ \justifying  Description of the noise model used in the simulation.}
\label{tab:noise}
\end{table}

\end{document}